\documentclass[twocolumn,showpacs,preprintnumbers,floats,aps]{revtex4}%
\usepackage{graphicx}
\usepackage{amsmath}
\usepackage{amsfonts}
\usepackage{amssymb}%
\begin{document}
\preprint{ }
\title{Nonlinear Dynamics in the Resonance Lineshape of NbN Superconducting Resonators}
\author{B. Abdo}
\email{baleegh@tx.technion.ac.il}
\author{E. Segev}
\author{O. Shtempluck}
\author{E. Buks}
\affiliation{Microelectronics Research Center, Department of Electrical Engineering,
Technion, Haifa 32000, Israel}
\date{\today}

\begin{abstract}
In this work we report on unusual nonlinear dynamics measured in the resonance
response of NbN superconducting microwave resonators. The nonlinear dynamics,
occurring at relatively low input powers (2-4 orders of magnitude lower than
Nb), and which include among others, jumps in the resonance lineshape,
hysteresis loops changing direction and resonance frequency shift, are
measured herein using varying input power, applied magnetic field, white noise
and rapid frequency sweeps. Based on these measurement results, we consider a
hypothesis according to which local heating of weak links forming at the
boundaries of the NbN grains are responsible for the observed behavior, and we
show that most of the experimental results are qualitatively consistent with
such hypothesis.

\end{abstract}
\maketitle




\section{INTRODUCTION}

Understanding the underlying mechanisms that cause and manifest nonlinear
effects in superconductors has a significant implications for both basic
science and technology. Nonlinear effects in superconductors may be exploited
to demonstrate some important quantum phenomena in the microwave regime as was
shown in Refs. \cite{squeezing,JPA}, and as was suggested recently in Refs.
\cite{Yurke,Eyal}. Whereas technologically, these effects can play a positive
or negative role depending on the application. On the one hand, they are very
useful in a wide range of nonlinear devices such as amplifiers \cite{rf
driven,Direct}, mixers \cite{mixer}, single photon detectors \cite{single
photon}, and SQUIDs \cite{flux qubit}. On the other hand, in other
applications mainly in the telecommunication area, such as band pass filters
and high Q resonators, nonlinearities are highly undesired \cite{Dahm,RF power
dependence,Intrinsic,nonlinear dynamics}.

Various nonlinear effects in superconductors and in NbN in particular have
been reported and analyzed in the past by several research groups. Duffing
like nonlinearity for example was observed in superconducting resonators
employing different geometries and materials. It was observed in a
high-$T_{c}$ superconducting (HTS)\ parallel plate resonator \cite{Thermally
induced nonlinear behaviour}, in a Nb microstrip resonator \cite{microwave
nonlinear effects in He-cooled sc microstrip resonators}, in a Nb and NbN
stripline resonators \cite{nonlinear dynamics}, in a YBCO coplanar-waveguide
resonator \cite{RF power dependence CP YBCO}, in a YBCO thin film dielectric
cavity \cite{Thermally-induced nonlinearities surface impedance sc YBCO}, and
also in a suspended HTS thin film resonator \cite{suspended HTS mw resonator}.
While other nonlinearities including notches, anomalies developing at the
resonance lineshape and frequency hysteresis were reported in Refs. \cite{HTS
patch antenna,power dependent effects observed for sc stripline resonator,mw
power handling weak links thermal effects}.

However, in spite of the intensive study of nonlinearities in superconductors
in the past decades, and the great progress achieved in this field, the
determination of the underlying mechanisms responsible for microwave nonlinear
behavior of both low- and high-$T_{c}$ superconductors is still a subject of
debate \cite{understanding}. This is partly because of the variety of
preparation and characterization techniques employed, and the numerous
fabrication parameters involved. In addition nonlinear mechanisms in
superconductors, which are usually divided into intrinsic and extrinsic, are
various and many times act concurrently. Thus identifying the dominant
mechanism is generally indirect \cite{Jerusalem}.

Among the nonlinear mechanisms investigated in superconductors one can name,
Meissner effect \cite{Yip}, pair-breaking, global and local heating effects
\cite{Thermally induced nonlinear behaviour,Thermally-induced nonlinearities
surface impedance sc YBCO}, rf and dc vortex penetration and motion
\cite{vortices}, defect points, damaged edges \cite{edge}, substrate material
\cite{anomalies in nonlinear mw surface vs sub effects}, and weak links (WL)
\cite{Halbritter}. Where WL is a collective term representing various material
defects such as, weak superconducting points switching to normal state under
low current density, Josephson junctions forming inside the superconductor
structure, grain-boundaries, voids, insulating oxides, insulating planes.
These defects and impurities generally affect the conduction properties of the
superconductor and as a result cause extrinsic nonlinear effects.

In this paper we report the observation of unique nonlinear effects measured
in the resonance lineshape of NbN superconducting microwave resonators. Among
the observed effects, asymmetric resonances, multiple jumps in the resonance
curve, hysteretic behavior in the vicinity of the jumps, frequency hysteresis
loops changing direction, jump frequency shift as the input power is increased
and nonlinear coupling. Some of these nonlinear effects were introduced by us
in a previous publication \cite{nonlinear features BB}. Thus this paper will
focus on presenting a new set of measurements applied to these nonlinear
resonators, which provides a better understanding of the underlying physical
mechanism causing these effects. To this end, we have measured the nonlinear
superconducting resonators using different operating conditions, such as
bidirectional frequency sweeps, added white noise, fast frequency sweep using
frequency modulation (FM), and dc magnetic field. In each case we observe a
unique nonlinear dynamics of the resonance lineshape which is qualitatively
different from the commonly reported Duffing oscillator nonlinearity. These
nonlinear effects are shown to originate from WL forming at the boundaries of
the NbN columnar structure. A theoretical model explaining the dynamical
behavior of the resonance lineshape in terms of abrupt changes in the
macroscopic parameters of the resonator is formulated. While these abrupt
transitions in the characteristic parameters of the resonator are attributed
to local heating of WL. Furthermore, simulations based on this model are shown
to be in a very good qualitative agreement with the experimental results.

The remainder of this paper is organized as follows, the fabrication process
of the NbN superconducting resonators is described briefly in Sec. II. The
nonlinear response of these resonators measured using various operating
conditions are reviewed in Sec. III. Comparison with other nonlinearities
reported in the literature is brought in Sec. IV. Possible underlying physical
mechanisms responsible for the observed effects are discussed in Sec. V.
Whereas in Sec. VI, a theoretical model based on local heating of weak links
is suggested, followed by simulations qualitatively reproducing most of the
nonlinear features observed in the experiments. Finally in Sec. VII, a short
summary concludes this paper.

\section{FABRICATION PROCESS}

\begin{figure}
[ptb]
\begin{center}
\includegraphics[
height=2.239in,
width=3.2699in
]%
{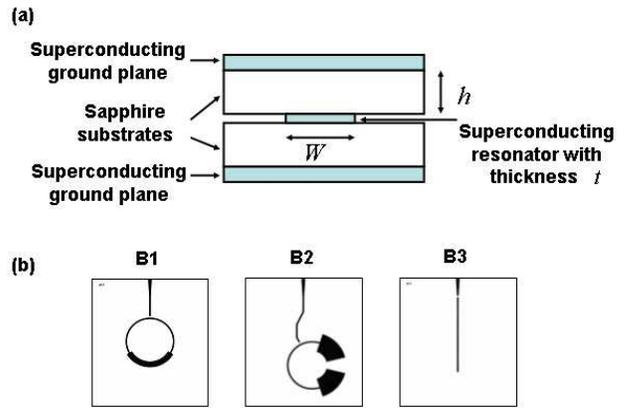}%
\caption{(a) Schematic cross section of the stripline geometry used,
consisting of five layers: two superconducting ground planes, two sapphire
substrates and a NbN film in the middle deposited on one of the sapphires. (b)
Top view of the three resonator layouts (B1, B2, B3) which were deposited as
the center layer. }%
\label{layout}%
\end{center}
\end{figure}

The measurement results presented in this paper belong to three nonlinear NbN
superconducting microwave resonators. The resonators were fabricated using
stripline geometry, consisting of two superconducting ground planes, two
sapphire substrates, and a center strip deposited in the middle (the
deposition was done on one of the sapphire substrates). Fig. \ref{layout}
shows a schematic diagram illustrating stripline geometry and a top view of
the three resonator-layouts. We will refer to the three resonators in the text
by the names B1, B2 and B3 as defined in Fig. \ref{layout}. The sapphire
substrates dimensions used were $34%
\mathrm{mm}%
$ X $30%
\mathrm{mm}%
$ X $1%
\mathrm{mm}%
,$ whereas the coupling gap between the resonators and their feedline was set
to $0.4%
\mathrm{mm}%
$ in B1 and B3 and $0.5%
\mathrm{mm}%
$ in B2 resonator. The resonators were dc-magnetron sputtered in a mixed
Ar/N$_{2}$ atmosphere, near room temperature. The patterning was done using
standard UV photolithography process, whereas the NbN etching was performed by
Ar ion-milling. The sputtering parameters, fabrication details, design
considerations as well as physical properties of the NbN films can be found
elsewhere \cite{nonlinear features BB}. The critical temperature $T_{c}$ of
B1, B2 and B3 resonators were relatively low and equal to $10.7%
\mathrm{K}%
,$ $6.8%
\mathrm{K}%
$ and $8.9%
\mathrm{K}%
$ respectively. The thickness of the NbN resonators were $2200$ $%
\mathrm{\text{\AA}}%
$ in B1, $3000$ $%
\mathrm{\text{\AA}}%
$ in B2, and $2000$ $%
\mathrm{\text{\AA}}%
$ in B3 resonator.%

\section{NONLINEAR RESONANCE RESPONSE}

In the following subsections, we present experimental results emphasizing the
different aspects of the nonlinear response exhibited by B1, B2 and B3
resonators. In subsection A, a resonance response measurement obtained while
varying the input rf power is presented, showing an abrupt and low power onset
of nonlinearity. In subsections B, C and D, representative experimental
results measured while scanning the resonance response in the forward and
backward directions are shown, exhibiting respectively, hysteresis loops
changing direction, metastability and multiple jumps. Whereas in subsection E,
the dependence of the resonance lineshape on applied dc magnetic field is
examined, showing a change in the direction of the jump and a jump vanishing
features. All measurements presented were performed at liquid helium
temperature $4.2%
\mathrm{K}%
$, and the results were verified using two configurations, immersing in liquid
helium and in vacuum.

\subsection{Abrupt onset of nonlinearity}

\begin{figure}
[ptb]
\begin{center}
\includegraphics[
height=2.4146in,
width=3.5924in
]%
{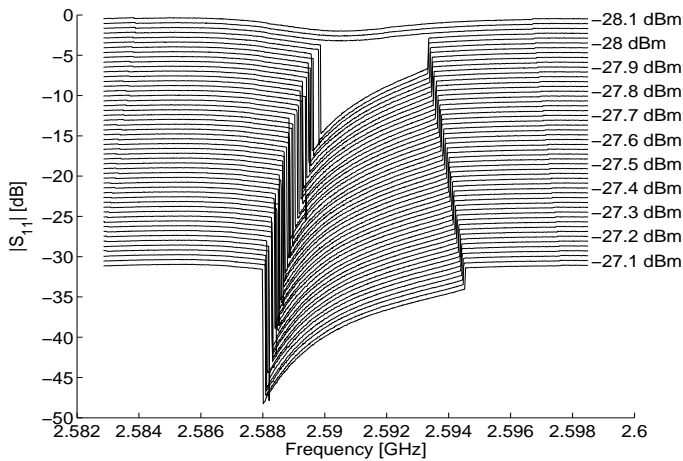}%
\caption{$S_{11}$ amplitude measurement of B1 resonator at its first mode. The
measured resonance lineshapes are asymmetrical and contain two abrupt jumps at
the sides of the resonance. Moreover the jump frequencies shift outwards as
the input power is increased. The measured resonance lineshapes were shifted
vertically by a constant offset for clarity.}%
\label{b1s11f1}%
\end{center}
\end{figure}

In Fig. \ref{b1s11f1} we present a $S_{11\text{ }}$parameter measurement of B1
first mode using a vector network analyzer. At low input powers, the resonance
response lineshape is Lorentzian and symmetrical. As the input power is
increased gradually in steps of $0.01$ dBm, the resonance response changes
dramatically and abruptly at about $-28.04$ dBm. It becomes extremely
asymmetrical, and includes two abrupt jumps at both sides of the resonance
lineshape. The magnitude of the jumps at some input powers, can be as high as
$16$ dB. As the input power is increased the resonance frequency is red
shifted and the jump frequencies shift outwards away from the center
frequency. Moreover at much higher powers not shown in the figure, the
resonance curve becomes gradually shallower and broader in the frequency span.
It is also worthwhile noting that the intensive evolution of the resonance
lineshape shown in Fig. \ref{b1s11f1} takes place within only $1$ dBm power range.%

\subsection{Hysteretic behavior}

As the equation of motion for nonlinear systems becomes multiple valued or
lacks a steady state solution in some parameter domain, nonlinear systems tend
to demonstrate hysteretic behavior with respect to that parameter.

Frequency hysteresis in the resonance lineshape of superconducting resonators
exhibiting Duffing oscillator nonlinearity and other kinds of nonlinearities
were observed by several groups \cite{HTS patch antenna,Nonlinear TL}.
Hysteretic behavior and losses in superconductors were discussed also in Refs.
\cite{vortices,vortex dynamics}. Moreover recent works, which examined the
resonance response of rf tank circuit coupled to a SQUID, have reported
several interesting frequency hysteresis features \cite{opposed
hammerhead,Pinch resonances in rf,nonlinear multilevel}.

\begin{figure}
[ptb]
\begin{center}
\includegraphics[
height=2.4189in,
width=3.5008in
]%
{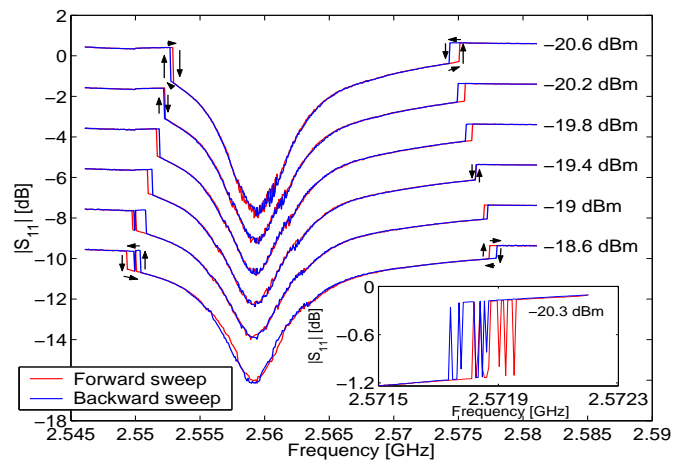}%
\caption{(Color online). Frequency sweep measurement of B1 resonator at its
first mode performed in both frequency directions. The plots exhibit
hysteresis loops forming at the vicinity of the jumps, as well as hysteresis
loops changing direction as the input power is increased. The red line
represents a forward sweep, whereas the blue line represents a backward sweep.
The number of measurement points employed in each scan direction is $500$
points. The resonance lineshapes were shifted vertically by a constant offset
for clarity. In the inset, a "Zoom in" measurement of the right hysteresis
loop of B1 first resonance is shown. The measurement, obtained using a
spectrum analyzer, includes $100$ data points and corresponds to $-20.3$ dBm
input power. }%
\label{hysteresis}%
\end{center}
\end{figure}

Likewise measuring the resonance response of our nonlinear resonators yields a
hysteretic behavior in the vicinity of the jumps. However, this hysteretic
behavior is unique in many aspects. In Fig. \ref{hysteresis} we show a
$\ S_{11\text{ }}$measurement of B1 resonator at its first mode, measured
while sweeping the frequency in both directions. The input power range shown
in this measurement corresponds to a higher power range than that of Fig.
\ref{b1s11f1}. The red line represents a forward frequency sweep, whereas the
blue line represents a backward frequency sweep.

At $-20.6$ dBm the resonance lineshape contains two jumps in each scan
direction and two hysteresis loops. The left hysteresis loop circulates
clockwise whereas the right loop circulates counter clockwise. However the
common property characterizing them is that the jumps occur at higher
frequencies in the forward scan compared to their counterparts in the backward
scan. As the input power is increased to about $-20.2$ dBm the two opposed
jumps at the left side meet and the left hysteresis loop vanishes. At about
$-19.4$ dBm a similar effect happens to the right hysteresis loop, and it
vanishes as well. Whereas at higher input powers (i.e. $-19$ dBm, $-18.6$ dBm)
the two jumps occur earlier at each frequency sweep direction, causing the
hysteresis loops to appear circulating in the opposite direction compared to
the $-20.6$ dBm resonance curve for instance. As we show in the next
subsection, this picture of well defined hysteresis loops is strongly
dependent on the applied frequency sweep rate and on the system noise. A
possible explanation for this unique hysteretic behavior would be presented in
Sec. VI.%

\subsection{Metastable states}

Jumps in the resonance response of a nonlinear oscillator are usually
described in terms of metastable/stable states and dynamic transition between
basins of attraction of the oscillator \cite{Cleland}, thus in order to
examine the stability of these observed resonance jumps, we carried out
several measurements.

In one measurement, we have "zoomed in" around the right jump of the resonance
at $-20.3$ dBm and examined its frequency response in both directions. The
measurement setup included a signal generator, the cooled resonator and a
spectrum analyzer. The reflected signal power off the resonator was redirected
by a circulator, and measured using a spectrum analyzer. The measurement
result obtained using $100$ sampling points in each direction, is exhibited in
the inset of Fig. \ref{hysteresis}, where the metastable nature of the jump
region is clearly demonstrated.

\begin{figure}
[ptbh]
\begin{center}
\includegraphics[
height=1.158in,
width=2.7717in
]%
{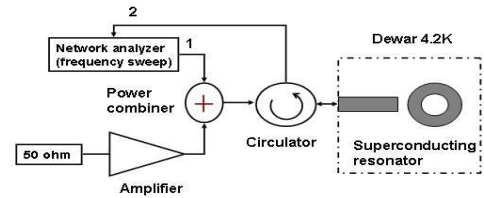}%
\caption{Schematic diagram of the experimental setup used to measure the
nonlinear first resonance of B1 using frequency sweep mode, while applying
white noise. }%
\label{noisesetup}%
\end{center}
\end{figure}

\begin{figure}
[ptbh]
\begin{center}
\includegraphics[
height=2.6316in,
width=3.563in
]%
{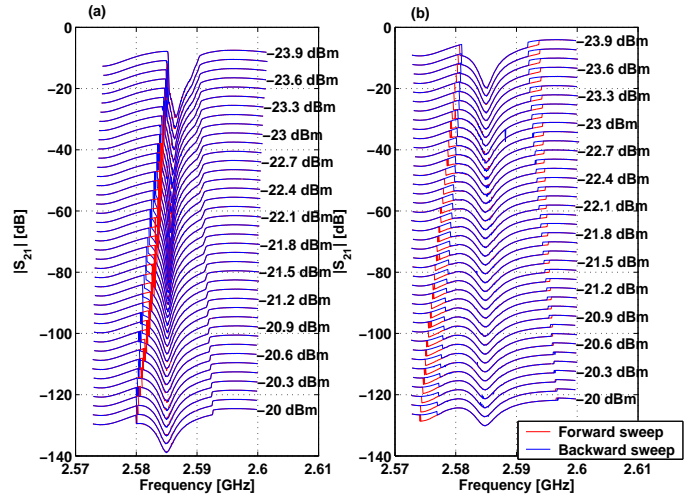}%
\caption{(Color online). Frequency sweep measurement of B1 resonator first
mode performed in both directions while (a) applying white noise of $-58$
dBm/Hz (b) without applying external noise. The red line represents a forward
sweep, whereas the blue line represents a backward sweep. The measured
resonance curves were shifted vertically by a constant offset for clarity.}%
\label{s21compnoise}%
\end{center}
\end{figure}

In another measurement configuration we have investigated this metastability
further by monitoring the effect of applied broadband noise on the resonance
jumps. We applied a constant white noise power to the resonator, several
orders of magnitude lower than the main signal power, using the setup depicted
in Fig. \ref{noisesetup}. The applied white noise level was $-58$ dBm/Hz
(measured separately using spectrum analyzer), and was generated by amplifying
the thermal noise of a room temperature $50$ $%
\mathrm{\Omega }%
$ load using an amplifying stage. The generated noise was added to the
transmitted power of a network analyzer via a power combiner. The power
reflections were redirected by a circulator, and were measured at the second
port of the network analyzer. The effect of the $-58$ dBm/Hz white noise power
on B1 first mode jumps, is shown in Fig. \ref{s21compnoise} (a), whereas in
Fig. \ref{s21compnoise} (b) we show for comparison the nearly noiseless case
obtained after disconnecting the amplifier and the combiner stage. The two
measurements were carried out in the same input power range ($-23.9$ dBm
through $-20$ dBm).

By comparing between the two measurement results, one can make the following
observations. The two fold jumps in Fig. \ref{s21compnoise} (b) form a
hysteresis loop at both sides of the resonance curve. By contrast in Fig.
\ref{s21compnoise} (a), as a result of the added noise, the hysteresis loops
at the right side vanish, while the jumps at the left side, become frequent
and bidirectional (indicated by the thick colored lines).

At a given input drive, the transition rate $\Gamma\left(  f\right)  $ between
the oscillator basins of attraction, can be generally estimated by the
expression $\Gamma\left(  f\right)  =\Gamma_{0}\exp\left(  -E_{A}\left(
f\right)  /k_{B}T_{eff}\right)  $ \cite{Cleland}$,$ where $E_{A}\left(
f\right)  $ is the quasi-activation energy of the oscillator, $T_{eff}$ is
proportional to the noise power, $k_{B}$ is Boltzmann's constant, $f$ is the
oscillator frequency, whereas $\Gamma_{0}$ is related to Kramers
low-dissipation form \cite{Kramer} and it is given approximately by $f_{0}/Q$,
where $f_{0}$ is the natural resonance frequency, and $Q$ is the quality
factor of the oscillator. From our results we roughly estimate the order of
magnitude of $E_{A}$ to be $10^{14}%
\mathrm{K}%
$ for the jump on the left \cite{Cleland}. Note however that this quantity
varies between different transitions and strongly depends on the operating point.%

%

\subsection{Multiple jumps}

\begin{figure}
[ptb]
\begin{center}
\includegraphics[
height=2.5382in,
width=3.3814in
]%
{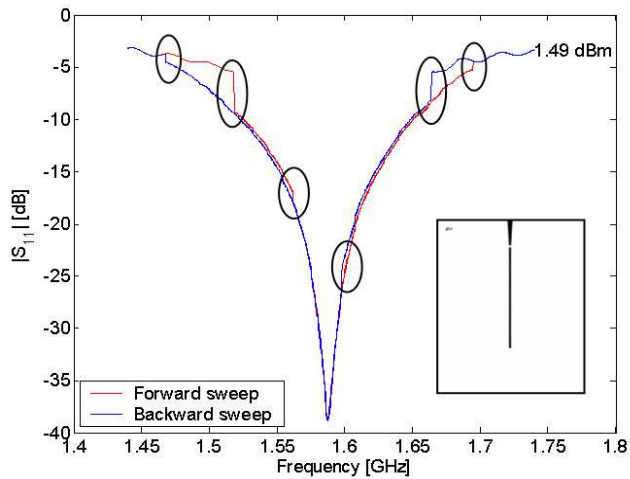}%
\caption{(Color online). $S_{11}$ parameter amplitude measurement of the first
resonance of B3 resonator, (shown in the inset), measured at input power of
$1.49$ dBm. The measurement was done using a network analyzer employing $4000$
measurement points in each direction. The red line represents a forward
frequency scan whereas the blue line represents a backward scan. The plot
shows clearly three jumps within the resonance lineshape in each direction, as
indicated by small circles. }%
\label{b3fb}%
\end{center}
\end{figure}

Another nonlinear feature, namely multiple jumps in the resonance lineshape,
is observed when measuring the resonance response of B3, while sweeping the
frequency in the forward and backward directions. In Fig. \ref{b3fb} we show a
representative measurement of the first resonance of B3 corresponding to
$1.49$ dBm input power, exhibiting three jumps in each sweep direction and
four hysteresis loops.%

\subsection{Dc magnetic field dependence}

\begin{figure}
[ptbh]
\begin{center}
\includegraphics[
height=2.3739in,
width=3.122in
]%
{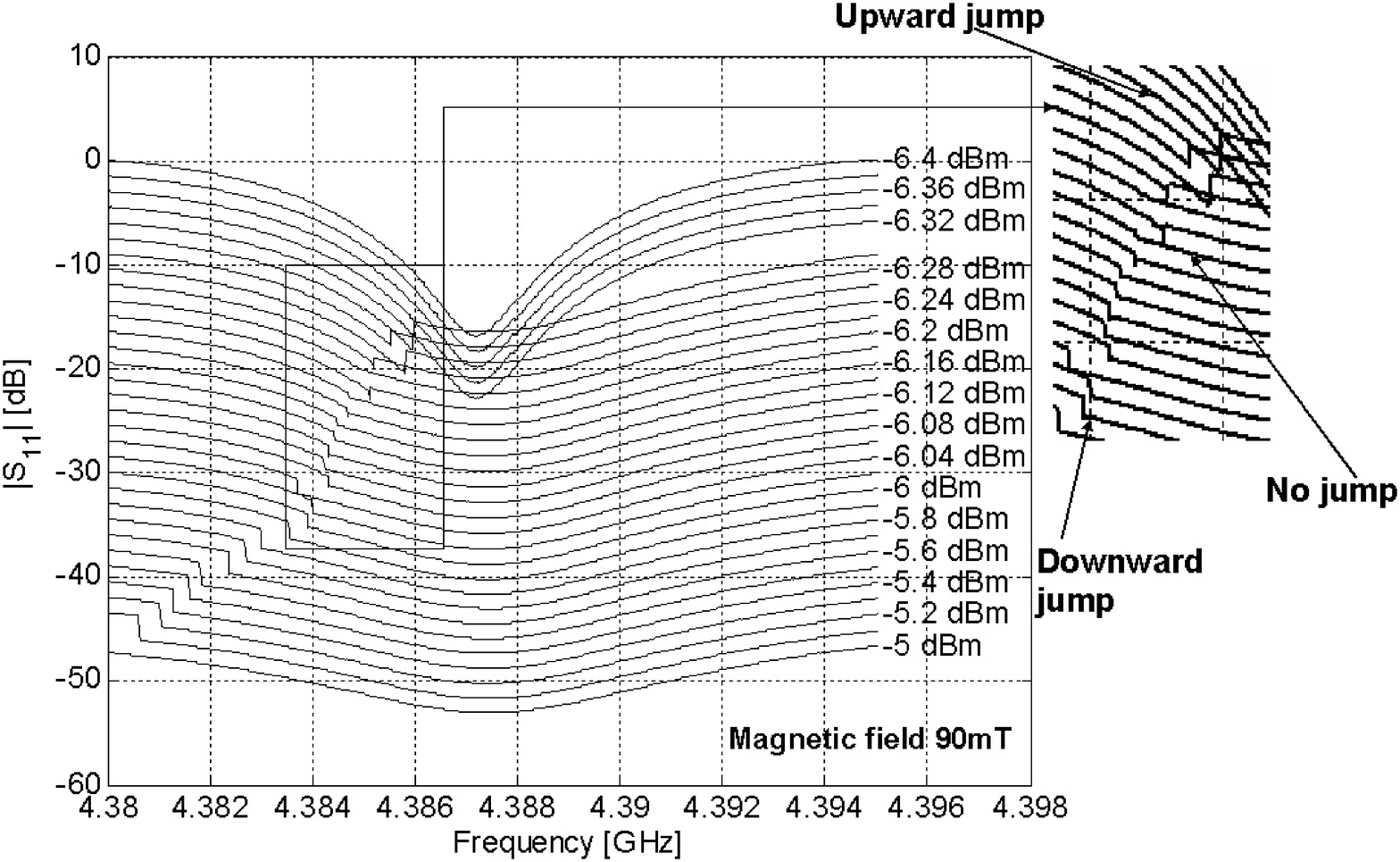}%
\caption{B2 nonlinear resonance response measured under a constant magnetic
field of $90\mathrm{mT},$ while increasing the input power. The
resonance which starts as a Lorentzian at low powers, develops into a
resonance curve having an upward jump, a curve with no jump, a curve having a
downward jump and finally a Lorentzian curve again as the power is increased.
The measured curves were shifted vertically by a constant offset for clarity.}%
\label{b2s11magconst}%
\end{center}
\end{figure}

\begin{figure}
[ptbh]
\begin{center}
\includegraphics[
height=2.0868in,
width=3.0182in
]%
{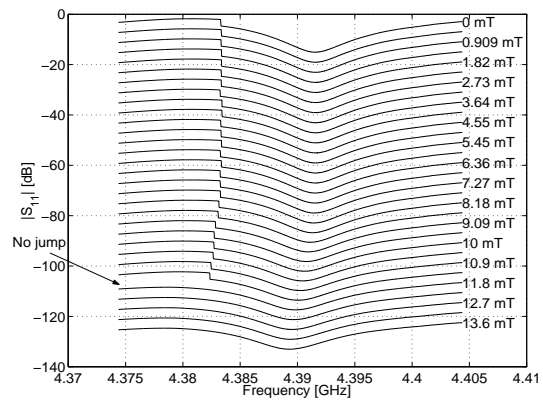}%
\caption{Increasing the magnetic field gradually from zero causes the jump in
the B2 resonance lineshape to disappear at relatively low value of
$11.8\mathrm{mT},$ while applying a constant input power level of $-5$
dBm. This jump vanishing indicates that the jump mechanism is sensitive to
magnetic field. The measured curves were shifted vertically by a constant
offset for clarity.}%
\label{magnojump}%
\end{center}
\end{figure}

Measuring B2 resonator second mode under dc magnetic field yielded additional
nonlinear features in the resonance response lineshape as shown in Figs.
\ref{b2s11magconst}, \ref{magnojump}.

In Fig. \ref{b2s11magconst} we show the resonance lineshape of B2 second mode
measured while applying a perpendicular dc magnetic field of $90%
\mathrm{mT}%
$. As the input rf power is increased gradually, the resonance lineshape
undergoes different phases. While at low and high powers the curves are
Lorentzians and symmetrical, in the intermediate range, the resonance curves
include a jump at the left side, which, as the input power increases, flips
from the upward to the downward direction.

Whereas in Fig. \ref{magnojump}, where we have set a constant input power of
$-5$ dBm and increased the applied magnetic field by small steps, the left
side jump vanishes as the magnetic field exceeds some relatively low threshold
of $\sim11.8%
\mathrm{mT}%
$. These effects will be further discussed in Secs. V and VI.%

%

\section{COMPARISON WITH OTHER NONLINEARITIES}

\begin{figure}
[ptb]
\begin{center}
\includegraphics[
height=2.2537in,
width=2.9888in
]%
{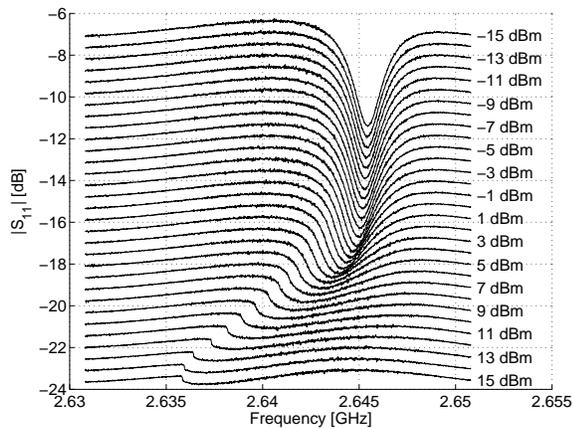}%
\caption{Duffing oscillator nonlinearity exhibited by a Nb stripline resonator
employing B2 layout at its first mode. The different $S_{11}$ amplitude plots
correspond to different input powers, ranging from $-15$ dBm to $15$ dBm in
steps of $1$ dBm. As the input power is increased the resonance becomes
asymmetrical and infinite slope builds up at the left side of the resonance
curve. The plots were offset in the vertical direction for clarity.}%
\label{nb_duffing}%
\end{center}
\end{figure}

The most commonly reported nonlinearity in superconductors is the Duffing
oscillator nonlinearity. However this nonlinearity is qualitatively different
from the nonlinearity we observe in our NbN samples and which is reported in
this paper. In Fig. \ref{nb_duffing} we show, for the sake of visual
comparison, a resonance response measured at $4.2%
\mathrm{K}%
$, exhibiting Duffing oscillator nonlinearity of the kind generally reported
in the literature \cite{nonlinear dynamics,Thermally induced nonlinear
behaviour,RF power dependence CP YBCO}. This nonlinearity which can be
explained in terms of resistance change $\Delta R$ and kinetic inductance
change $\Delta L_{K}$ \cite{Dahm,Nonlinear TL} was measured at the first
resonance frequency of a $2200$ $%
\mathrm{\text{\AA}}%
$ thickness Nb resonator employing B2 layout geometry ($T_{c}=8.9%
\mathrm{K}%
$). The differences between the two nonlinear dynamics shown in Figs.
\ref{b1s11f1} and \ref{nb_duffing}, are obvious. In Fig. \ref{nb_duffing} the
nonlinearity is gradual, while in Fig. \ref{b1s11f1} the power onset of
nonlinearity is abrupt and sudden. In Fig. \ref{nb_duffing}, the resonance
response in the nonlinear regime, contains an infinite slope at the left side,
whereas in Fig. \ref{b1s11f1} the curves contain two jumps at both sides of
the resonance response. In Fig. \ref{nb_duffing} changes in the resonance
curve are measured on power scale of $1$ dBm, while changes in Fig.
\ref{b1s11f1} are measured on $0.01$ dBm scale. Whereas the onset of
nonlinearity in Fig. \ref{nb_duffing} is of the order of $10$ dBm, the onset
of nonlinearity in Fig. \ref{b1s11f1} is about $4$ orders of magnitudes lower
$\sim-28$ dBm. Furthermore, by taking into account the differences in the
hysteretic behavior of the two nonlinearities and the multiple jumps feature
shown in Fig. \ref{b3fb}, the special characteristics of the reported
nonlinearity are further established.

Abrupt jumps in the resonance lineshape similar in some aspects to the jumps
reported herein, were observed in two-port high-$T_{c}$ YBCO resonators
\cite{HTS patch antenna,power dependent effects observed for sc stripline
resonator,mw power handling weak links thermal effects}. Portes \textit{et
al.} \cite{HTS patch antenna} have also reported some frequency hysteretic
behavior in the vicinity of the jumps. However, one significant difference
between the two nonlinearities is the onset power of nonlinearity reported in
these references, which is on the order of $20$ dBm \cite{power dependent
effects observed for sc stripline resonator,mw power handling weak links
thermal effects}, that is about $5$ orders of magnitude higher than the onset
power of nonlinearity of B1 first mode ($\sim-28$ dBm). All three works
\cite{HTS patch antenna,power dependent effects observed for sc stripline
resonator,mw power handling weak links thermal effects} have attributed the
nonlinear abrupt jumps to local heating of distributed WL in the resonator film.%

\section{POSSIBLE NONLINEAR MECHANISMS}

The relatively very low onset power of nonlinearity observed in these
resonators as well as its strong sensitivity to rf power, highly imply an
extrinsic origin of these effects, and as such, hot spots in WL is a leading
candidate for explaining the nonlinearity.

Vortex penetration in the bulk or WL is less likely, mainly because heating
the sample above $T_{c}$ between sequential magnetic field measurements, have
yielded reproducible results with a good accuracy in the magnetic field
magnitude, the microwave input power and in the jump frequency (less than
$200$ $%
\mathrm{kHz}%
$ offset). Moreover the low magnetic field threshold $\sim11.8$ $%
\mathrm{mT}%
$ above which the B2 resonance jump vanished, is about $3.5$ times lower than
$H_{c_{1}}$ (flux penetration) of NbN reported for example in \cite{nonlinear
dynamics}.

In the following subsection A, we provide a direct evidence of WL. Whereas in
subsection B, we exclude global heating mechanism as a possible source of the effects.

\subsection{Columnar structure}

\begin{figure}
[ptbh]
\begin{center}
\includegraphics[
height=2.2148in,
width=3.3944in
]%
{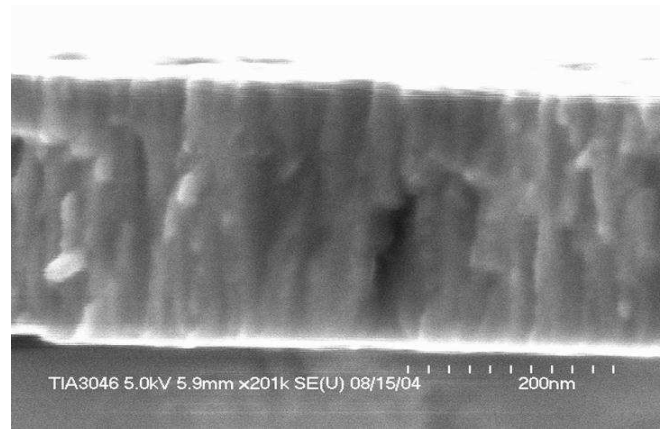}%
\caption{A SEM micrograph displaying a $2200\mathrm{\text{\AA}}$ NbN
film deposited on a thin sapphire substrate using similar sputtering
conditions as B2 resonator. The micrograph exhibits clearly the columnar
structure of the NbN film and its grain boundaries.}%
\label{sem}%
\end{center}
\end{figure}

It is well known from numerous research works done in the past \cite{rf
superconducting properties of reactively sputtered NbN,superconducting
properties and NbN structure}, that NbN films can grow in a granular columnar
structure under certain deposition conditions. Such columnar structure may
even promote the growth of random WL at the grain boundaries of the NbN films.
To study the NbN granular structure we have sputtered about $2200%
\mathrm{\text{\AA}}%
$ NbN film on a thin small rectangular sapphire substrate of $0.2$ $%
\mathrm{mm}%
$ thickness. The sputtering conditions applied were similar to those used in
the fabrication of B2 resonator. Following the sputtering process, the thin
sapphire was cleaved, and a scanning electron microscope (SEM) micrograph was
taken at the cleavage plane. The SEM micrograph in Fig. \ref{sem} which
clearly shows the columnar structure of the deposited NbN film and its grain
boundaries, further supports our weak link hypothesis. The typical diameter of
each NbN column is about $20$ $%
\mathrm{nm}%
.$%

\subsection{Frequency sweep time analysis}

\begin{figure}
[ptbh]
\begin{center}
\includegraphics[
height=1.689in,
width=3.4324in
]%
{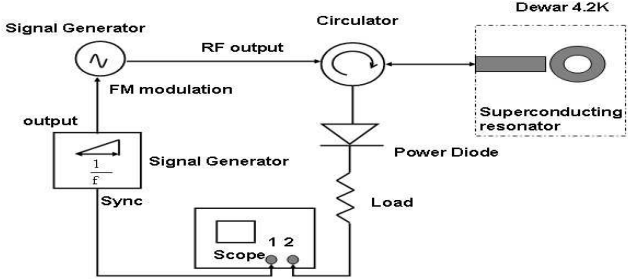}%
\caption{Frequency sweep time analysis setup. The frequency sweep time of the
microwave signal generator was FM modulated by a saw tooth waveform with
frequency $f$. The reflected power from the resonator was measured by a power
diode and oscilloscope.}%
\label{fmsetup}%
\end{center}
\end{figure}

\begin{figure}
[ptbh]
\begin{center}
\includegraphics[
height=2.0825in,
width=3.4065in
]%
{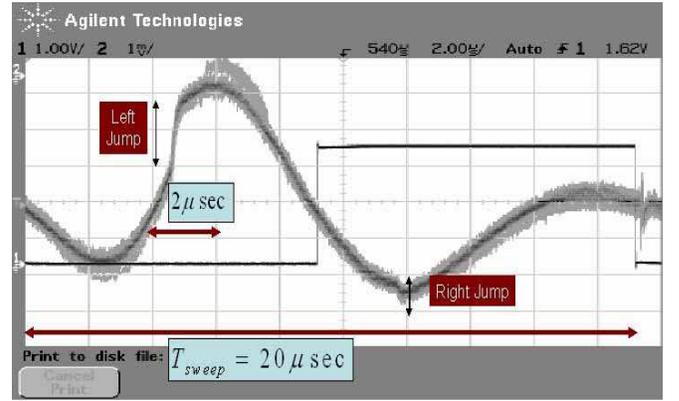}%
\caption{Frequency sweep time measurement. The figure displays the resonance
measured by Agilent oscilloscope while applying a saw tooth FM modulation of
frequency $50\mathrm{kHz}$ ( $T_{sweep}=$20$\mathrm{\mu s}$) to
the signal generator. The left and right jumps of the resonance are still
apparent in spite of the fast rate frequency sweep. Thus indicating that the
jumps do not originate from any global heating mechanism.}%
\label{timesweep}%
\end{center}
\end{figure}

Resistive losses and heating effects are typically characterized by relatively
long time scales \cite{Thermally induced nonlinear behaviour}. In attempt to
consider whether such effects are responsible for the observed nonlinearities
in general and for the jumps in particular, we have run frequency sweep time
analysis using the experimental setup depicted in Fig. \ref{fmsetup}. We have
controlled the frequency sweep cycle of a signal generator via FM modulation.
The FM modulation was obtained by feeding the signal generator with a saw
tooth waveform having $1/f$ sweep time cycle. The reflected power from the
resonator was redirected using a circulator and measured by a power diode and
oscilloscope. The left and right hand jumps of B2 $\sim4.39%
\mathrm{GHz}%
$ resonance were measured using this setup, while applying increasing FM
modulation frequencies up to $200%
\mathrm{kHz}%
$. In Fig. \ref{timesweep} we present a measurement result obtained at $50%
\mathrm{kHz}%
$ FM modulation, or alternatively $T_{sweep}$ of $20%
\mathrm{\mu s}%
$. The FM modulation applied was $\pm20%
\mathrm{MHz}%
$ around $4.4022%
\mathrm{GHz}%
$ center frequency. The measured resonance response appears inverted in the
figure due to the negative output polarity of the power diode. The fact that
both jumps continue to occur within the resonance lineshape (see Fig.
\ref{timesweep}), in spite of the short duty cycles that are of the order of
$\sim%
\mathrm{\mu s}%
$, indicates that heating processes which have typical time scale on the order
of $%
\mathrm{s}%
$ to $%
\mathrm{ms}%
$ \cite{Thermally induced nonlinear behaviour} are unlikely to cause these effects.%

%

However the above measurement result does not exclude local heating of WL
\cite{Bol,Nonequilibrium,Use}. Assuming that the substrate is isothermal and
that the hot spot is dissipated mainly down into the substrate rather than
along the film \cite{Bol}, one can evaluate the characteristic relaxation time
of the hot spot using the equation $\tau=Cd/\alpha,$ where $C$ is the heat
capacity of the superconducting film (per unit volume), $d$ is the film
thickness, and $\alpha$ is the thermal surface conductance between the film
and the substrate \cite{Nonequilibrium}. Substituting for our B2 NbN resonator
yields a characteristic relaxation time of $\tau\simeq5.4\cdot10^{-8}%
\mathrm{s}%
,$ where the parameters $C\simeq2.7\cdot10^{-3}%
\mathrm{J}%
\mathrm{cm}%
^{-3}%
\mathrm{K}%
^{-1}$ (NbN) \cite{Use}, $d=3000%
\mathrm{\text{\AA}}%
$ (B2 thickness)$,$ and $\alpha\simeq1.5%
\mathrm{W}%
\mathrm{cm}%
^{-2}%
\mathrm{K}%
^{-1}$ at $4.2%
\mathrm{K}%
$ (sapphire substrate) \cite{Use}, have been used. Similar calculation based
on values given in Ref. \cite{Bol} yields $\tau\simeq2.1\cdot10^{-9}%
\mathrm{s}%
.$ These time scales are of course 2-3 orders of magnitude lower than the time
scales examined by the FM modulation setup, and thus local heating of WL is
not ruled out.

\section{LOCAL HEATING of WL MODEL}

In this section we consider a hypothesis according to which local heating of
WL is responsible for the observed effects. We show that this hypothesis can
account for the main nonlinear features observed, and that simulations based
on such a theoretical model, exhibit a very good qualitative agreement with
experimental results.

\subsection{Theoretical modeling}

Consider a resonator driven by a weakly coupled feedline carrying an incident
coherent tone $b^{in}e^{-i\omega_{p}t}$, where $b^{in}$ is a constant complex
amplitude and $\omega_{p}$ is the drive angular frequency. The mode amplitude
inside the resonator $A$ can be written as $A=Be^{-i\omega_{p}t}$, where
$B\left(  t\right)  $ is a complex amplitude, which is assumed to vary slowly
on the time scale of $1/\omega_{p}$. In this approximation, the equation of
motion of $B$ reads \cite{Yurke}%

\begin{equation}
\frac{dB}{dt}=\left[  i\left(  \omega_{p}-\omega_{0}\right)  -\gamma\right]
B-i\sqrt{2\gamma_{1}}b^{in}+c^{in}, \label{dB/dt}%
\end{equation}

where $\omega_{0}$ is the angular resonance frequency, $\gamma=\gamma
_{1}+\gamma_{2}$, $\gamma_{1}$ is the coupling constant between the resonator
and the feedline, and $\gamma_{2}$ is the damping rate of the mode. The term
$c^{in}$ represents input noise with vanishing average%

\begin{equation}
\langle c^{in}\rangle=0,
\end{equation}

and correlation function given by%

\begin{equation}
\langle c^{in}(t)c^{in\ast}(t^{\prime})\rangle=G\omega_{0}\delta\left(
t-t^{\prime}\right)  . \label{corr}%
\end{equation}

In thermal equilibrium and for the case of high temperature $k_{B}%
T>>\hbar\omega_{0}$, where $k_{B}$ is Boltzmann's constant, one has%

\begin{equation}
G=\frac{2\gamma}{\omega_{0}}\frac{k_{B}T}{\hbar\omega_{0}}.
\end{equation}
\ 

In terms of the dimensionless time $\tau=\omega_{0}t,$ Eq. (\ref{dB/dt}) reads%

\begin{equation}
\frac{dB}{d\tau}=\frac{i\left(  \omega_{p}-\omega_{0}\right)  -\gamma}%
{\omega_{0}}\left(  B-B_{\infty}\right)  +\frac{c^{in}}{\omega_{0}},
\label{dB/d_tau}%
\end{equation}

where%

\begin{equation}
B_{\infty}=\frac{i\sqrt{2\gamma_{1}}b^{in}}{i\left(  \omega_{p}-\omega
_{0}\right)  -\gamma}. \label{Binf}%
\end{equation}

Small noise gives rise to fluctuations around the steady state solution
$B_{\infty}$. A straightforward calculation yields%

\begin{equation}
\left\langle \left\vert B-B_{\infty}\right\vert ^{2}\right\rangle
=\frac{G\omega_{0}}{2\gamma}.
\end{equation}
The output signal $a^{out}$ reflected off the resonator can be written as
$a^{out}=b^{out}e^{-i\omega_{p}t}$. The input-output relation relating the
output signal to the input signal is given by \cite{Gardiner}%

\begin{equation}
\frac{b^{out}}{\sqrt{\omega_{0}}}=\frac{b^{in}}{\sqrt{\omega_{0}}}%
-i\sqrt{\frac{2\gamma_{1}}{\omega_{0}}}B. \label{in_out}%
\end{equation}

Whereas the total power dissipated in the resonator $Q_{t}$ can be expressed
as \cite{Yurke}%

\begin{equation}
Q_{t}=\hslash\omega_{0}2\gamma_{2}E, \label{Qt}%
\end{equation}

where $E=\left\vert B\right\vert ^{2}$.

Furthermore, consider the case where the nonlinearity is originated by a local
hot spot in the stripline resonator. If the hot spot is assumed to be
sufficiently small, its temperature $T$ can be considered homogeneous. The
temperature of other parts of the resonator is assumed be equal to that of the
coolant $T_{0}$. The power $Q$ heating up the hot spot is given by $Q=\alpha
Q_{t}$ where $0\leqslant\alpha\leqslant1$.

The heat balance equation reads%

\begin{equation}
C\frac{dT}{dt}=Q-W, \label{heat_balance}%
\end{equation}

where $C$ is the thermal heat capacity, $W=H\left(  T-T_{0}\right)  $ is the
power of heat transfer to the coolant, and $H$ the heat transfer coefficient.
Defining the dimensionless temperature \cite{hot spots}%

\begin{equation}
\Theta=\frac{T-T_{0}}{T_{c}-T_{0}}, \label{Theta}%
\end{equation}

where $T_{c}$ is the critical temperature, one has%

\begin{equation}
\frac{d\Theta}{d\tau}=-g\left(  \Theta-\Theta_{\infty}\right)  ,
\label{dT/d_tau}%
\end{equation}

where%

\begin{equation}
\Theta_{\infty}=\frac{2\alpha\gamma_{2}\rho E}{\omega_{0}g}, \label{Theta_inf}%
\end{equation}

\begin{equation}
\rho=\frac{\hslash\omega_{0}}{C\left(  T_{c}-T_{0}\right)  }, \label{rho}%
\end{equation}

\begin{equation}
g=\frac{H}{C\omega_{0}}. \label{g}%
\end{equation}

While in Duffing oscillator equation discussed in Ref. \cite{Yurke}, the
nonlinearity can be described in terms of a gradually varying resonance
frequency dependent on the amplitude of the oscillations inside the cavity, in
the current case, the resonance frequency $\omega_{0}$, the damping rates
$\gamma_{1}$, $\gamma_{2}$ and $\alpha$ factor are considered to have a step
function dependence on $T$, the temperature of the WL%

\begin{equation}
\omega_{0}=\left\{
\begin{array}
[c]{cc}%
\omega_{0s} & \Theta<1\\
\omega_{0n} & \Theta>1
\end{array}
\right.  , \label{w0}%
\end{equation}

\begin{equation}
\gamma_{1}=\left\{
\begin{array}
[c]{cc}%
\gamma_{1s} & \Theta<1\\
\gamma_{1n} & \Theta>1
\end{array}
\right.  , \label{g1}%
\end{equation}

\begin{equation}
\gamma_{2}=\left\{
\begin{array}
[c]{cc}%
\gamma_{2s} & \Theta<1\\
\gamma_{2n} & \Theta>1
\end{array}
\right.  , \label{g2}%
\end{equation}

\begin{equation}
\alpha=\left\{
\begin{array}
[c]{cc}%
\alpha_{s} & \Theta<1\\
\alpha_{n} & \Theta>1
\end{array}
\right.  . \label{alpha}%
\end{equation}

In general, while disregarding noise, the coupled differential equations
(\ref{dB/d_tau}) and (\ref{dT/d_tau}) may have up to two different steady
state solutions. A superconducting steady state of the WL exists when
$\Theta_{\infty}<1$, or alternatively when $E<E_{s}$, where $E_{s}=gC\left(
T_{c}-T_{0}\right)  /2\alpha_{s}\gamma_{2s}\hslash$. Similarly, a normal
steady state of the WL exists when $\Theta_{\infty}>1$, or alternatively when
$E>E_{n}$, where $E_{n}=gC\left(  T_{c}-T_{0}\right)  /2\alpha_{n}\gamma
_{2n}\hslash$.

In general, the reflection coefficient $S_{11}$ in steady state is given by
\cite{Yurke}%

\begin{equation}
S_{11}=\frac{b^{out}}{b^{in}}=\frac{\gamma_{2}-\gamma_{1}-i\left(  \omega
_{p}-\omega_{0}\right)  }{\gamma_{2}+\gamma_{1}-i\left(  \omega_{p}-\omega
_{0}\right)  }. \label{r}%
\end{equation}

\subsection{Simulation results}

Simulating the resonator system using this local heating WL model, yields
results which qualitatively agree with most of the nonlinear effects
previously presented. In subsection (1) we simulate the main effects of Sec.
III (A, B, C, D), whereas in subsection (2) we simulate and provide a possible
explanation to the nonlinear features of Sec. III (E).

\subsubsection{Abrupt jumps and hysteretic behavior}

\begin{figure}
[ptb]
\begin{center}
\includegraphics[
height=2.4647in,
width=2.9741in
]%
{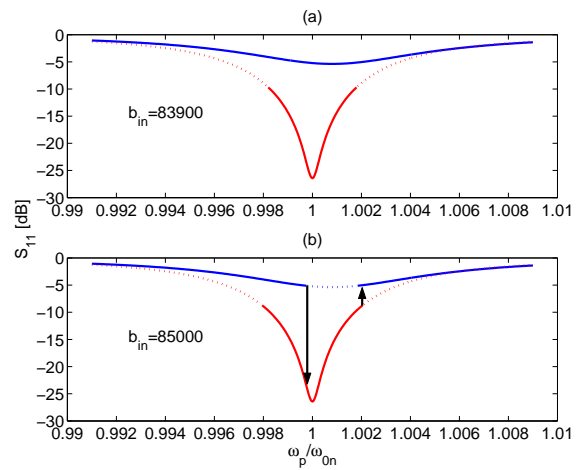}%
\caption{(Color online). Simulated resonance response obtained by the hot spot
mechanism modeling. The different plots simulate the nonlinear behavior shown
in Fig. \ref{b1s11f1}. Panels (a) and (b) correspond to an increasing drive
amplitude $b^{in}$. The solid lines represent valid steady state solutions
whereas the dotted lines represent invalid solutions. The red lines represent
the normal WL solutions, whereas the blue lines represent the superconducting
WL solutions. The black arrows show the direction of the jumps in the
different cases. The parameters that were used in the simulation are:
$\omega_{0s}/\omega_{0n}=1.0008,$ $\gamma_{1n}/\omega_{0n}=2.5\cdot10^{-3},$
$\gamma_{1s}/\omega_{0n}=1.5\cdot10^{-3},$ $\gamma_{2n}/\omega_{0n}%
=2.75\cdot10^{-3},$ $\gamma_{2s}/\omega_{0n}=5\cdot10^{-3},$ $\alpha_{n}=0.8,$
$\alpha_{s}=1,$ $g=0.5,$ $\rho=10^{-10}$. }%
\label{two_wells}%
\end{center}
\end{figure}

In Fig. \ref{two_wells} we show a resonance response simulation result based
on the hot spot model, which simulates the abrupt jump exhibited in Fig.
\ref{b1s11f1}. The solid and dotted lines represent valid steady state
solutions of the system and invalid steady state solutions respectively.
Whereas the blue and red colors represent superconducting WL solutions and
normal WL solutions respectively. In plot (a) the superconducting WL solution
is valid in the normalized frequency span, and therefore the system follows
this lineshape without jumps. As we increase the amplitude drive $b_{in}$ we
obtain a result seen in plot (b). As the frequency is swept, jumps in the
resonance response are expected to take place as the solution followed by the
system (according to the initial conditions) becomes invalid. Thus in the
forward sweep direction (as the frequency sweep of Fig. \ref{b1s11f1}), we get
two jumps indicated by black arrows on the figure. Similar to Fig.
\ref{b1s11f1}, the magnitudes of the jumps in the plot are unequal (the left
jump is higher). This difference in the magnitude of the jumps is generally
dependent on the relative position between the two resonance frequencies (Eq.
\ref{w0}), while in measurement due to the metastability of the system in the
hysteretic regime, it depends also on the frequency sweep rate. The simulation
parameters used in the different cases are indicated in the figure captions.%

\begin{figure}
[ptb]
\begin{center}
\includegraphics[
height=2.4535in,
width=3.0493in
]%
{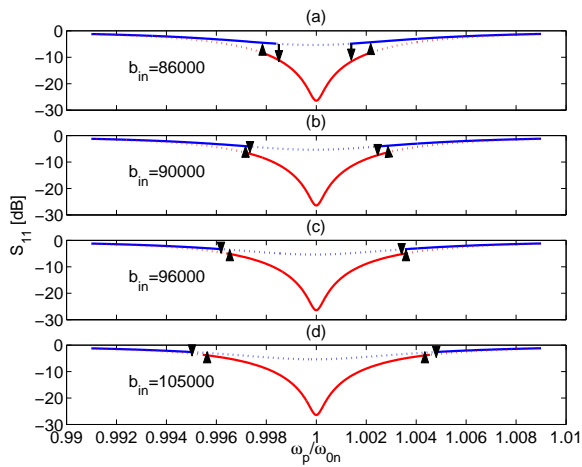}%
\caption{(Color online). Simulated resonance response obtained by the hot spot
mechanism modeling. Panels (a), (b), (c) and (d) correspond to an increasing
drive amplitude $b^{in}$. The different plots simulate the nonlinear behavior
shown in Fig. \ref{hysteresis}. The lines and symbols are the same as in Fig.
\ref{two_wells}. The parameters that were used in the simulation are:
$\omega_{0s}/\omega_{0n}=0.99989,$ $\gamma_{1n}/\omega_{0n}=2.5\cdot10^{-3},$
$\gamma_{1s}/\omega_{0n}=1.5\cdot10^{-3},$ $\gamma_{2n}/\omega_{0n}%
=2.75\cdot10^{-3},$ $\gamma_{2s}/\omega_{0n}=5\cdot10^{-3},$ $\alpha_{n}=0.8,$
$\alpha_{s}=1,$ $g=0.5,$ $\rho=10^{-10}$. }%
\label{twodirmod}%
\end{center}
\end{figure}

The behavior of the frequency hysteresis loops exhibited in Fig.
\ref{hysteresis} is simulated in Fig. \ref{twodirmod}. The different plots
exhibited in Fig. \ref{twodirmod} correspond to the different cases shown in
Fig. \ref{hysteresis}. In plot (a) the jumps in the forward direction
(indicated by the arrows in that direction) occur at higher frequencies than
the jumps in the backward direction. Whereas in plot (b) corresponding to a
higher amplitude drive $b_{in}$ we show a case in which the left side
hysteresis loop vanish as the two opposed jump frequencies coincide. If we
increase $b_{in}$ further, then at some amplitude drive level as shown in plot
(c), we get a similar case of hysteresis loop vanishing at the right side of
the resonance response. Whereas at the left side we get a frequency region
where both the superconducting and the normal WL solutions are invalid. In
this instable region, transitions between the invalid solutions are expected,
depending on the number of the sampling points, the sweep time, and the
internal noise. However due to this instability, the system is highly expected
to jump "early" in each frequency direction, as it enters this region (at
lower frequencies in the forward direction, and at higher frequencies in the
backward direction). Thus leading to the observed change in the direction of
the hysteresis loop. By increasing $b_{in}$ further, one obtains a case in
which both hysteresis loops are circulating in the opposite direction compared
to plot (a).%

\begin{figure}
[ptb]
\begin{center}
\includegraphics[
height=1.9527in,
width=2.8807in
]%
{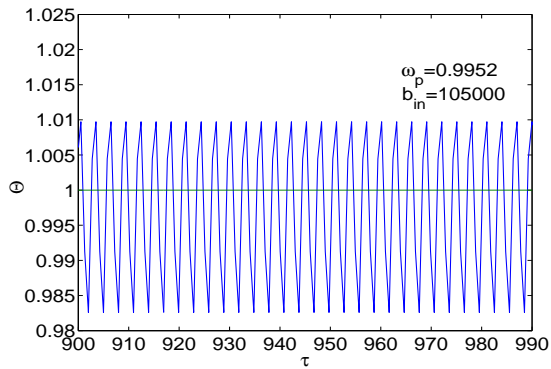}%
\caption{The dynamical solution of the coupled Eqs. \ref{dB/d_tau},
\ref{dT/d_tau} at a normalized frequency of $\omega_{p}=0.9952$. The
simulation parameters used are the same as those of plot (d) in Fig.
\ref{twodirmod}. }%
\label{thetatau}%
\end{center}
\end{figure}

Furthermore the intermediate jump indicating instability, which appear at the
left jump region of the last resonance curve in Fig. \ref{hysteresis}
(corresponding to $-18.6$ dBm), can be explained by this model as well. By
solving the coupled Eqs. (\ref{dB/d_tau}), (\ref{dT/d_tau}) in the time domain
for a single normalized frequency $\omega_{p}=0.9952$ (arbitrarily chosen in
the left side hysteresis region) and using the simulation parameters of plot
(d) in Fig. \ref{twodirmod}, one obtains the oscillation pattern of the
dimensionless parameter $\Theta$ shown in Fig. \ref{thetatau}, as a function
of the dimensionless time $\tau$. The $\Theta$ oscillations indicating
instability are between the superconducting and the normal values,
corresponding to $\Theta<1$ and $\Theta>1$ respectively.%

As to the multiple jumps feature exhibited in Fig. \ref{b3fb}, a
straightforward generalization of the model may be needed in order to account
for this effect. Such generalization would include WL having a variation in
their sizes and critical current along the stripline. Thus causing them to
switch to normal state at different current drives (at different frequencies)
and as a result induce more than two jumps in the resonance lineshape.

\subsubsection{Magnetic field dependence}

In this subsection we show how the model of local heating of WL can also
account for the nonlinear dynamics of the resonance lineshape observed under
applied magnetic field.

\begin{figure}
[ptb]
\begin{center}
\includegraphics[
height=2.444in,
width=3.1756in
]%
{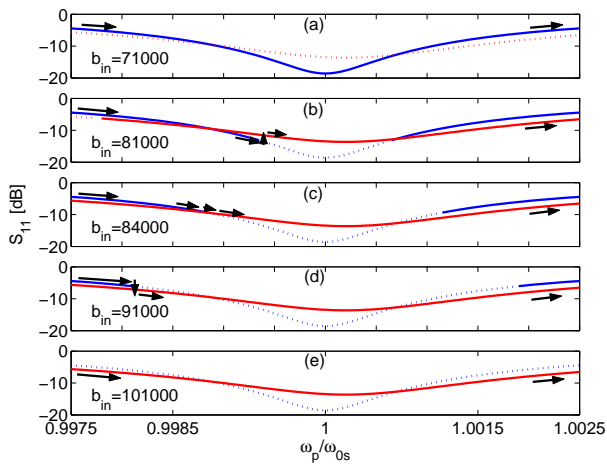}%
\caption{(Color online). Simulated resonance response obtained by the hot spot
mechanism modeling. The different plots simulate the nonlinear behavior shown
in Fig. \ref{b2s11magconst}. Panels (a), (b), (c), (d) and (e) correspond to
an increasing drive amplitude $b^{in}$. The lines and symbols are the same as
in Fig. \ref{two_wells}. The simulation parameters used are: $\omega
_{0n}/\omega_{0s}=1.0002,$ $\gamma_{1n}/\omega_{0s}=0.0029,$ $\gamma
_{1s}/\omega_{0s}=0.0019,$ $\gamma_{2n}/\omega_{0s}=0.0019,$ $\gamma
_{2s}/\omega_{0s}=0.0015,$ $\alpha_{n}=1,$ $\alpha_{s}=0.8,$ $g=0.5,$
$\rho=10^{-10}$. }%
\label{magsim}%
\end{center}
\end{figure}

To this end, we show in Fig. \ref{magsim} a simulation result based on the WL
local heating model, which qualitatively regenerates the nonlinear behavior of
the resonance lineshape of B2 under a constant magnetic field (presented in
Fig. \ref{b2s11magconst}). At low drive amplitude $b_{in}$, only the
superconducting WL steady state solution exists, and thus no jump occurs as
one sweeps the frequency (plot (a)). Increasing the drive amplitude $b_{in}$
(plot (b)) causes the superconducting WL solution to become invalid in the
center frequency region, thus the resonance response jumps upward (as the
system reaches the invalid region) and stabilizes on the normal WL steady
state solution, as indicated by arrows on the plot. By increasing the drive
amplitude further (plot (c)) there exists an intersection point where a smooth
transition without a jump is expected to occur between the valid
superconducting WL solution and the valid normal one. Whereas in plot (d)
where we have increased $b_{in}$ further, a downward jump in the resonance
response occurs as the valid normal WL solution lies below the invalid
superconducting WL solution. Finally in plot (e) corresponding to a much
higher drive, only the normal WL steady state solution exists within the
frequency span and therefore there are no jumps in the resultant curve.%

Another measurement which can be explained using the WL model, is the
measurement shown in Fig. \ref{magnojump}, where the left side jump vanishes
as the magnetic field increases above some low magnetic field threshold. This
result can be explained in the following manner. Increasing the applied dc
magnetic field would inevitably increase the screening supercurrent flowing in
the film and the local heating of WL. As the local heating exceeds some
threshold, the superconducting WL solution would become invalid (for the same
frequency span), and consequently the system would only follow the normal WL
solution without apparent jumps.

\section{SUMMARY}

In attempt to investigate and manifest nonlinear effects in superconducting
microwave resonators, several superconducting NbN resonators employing
different layouts, but similar sputtering conditions have been designed and
fabricated. The resonance lineshapes of these NbN resonators having very low
onset of nonlinearity, several orders of magnitude lower than other reported
nonlinearities \cite{nonlinear dynamics,mw hysteretic losses in YBCO and
NbN,Palomba}, exhibit some extraordinary nonlinear dynamics. Among the
nonlinearities observed, while applying different measurement configurations,
abrupt metastable jumps in the resonance lineshape, hysteresis loops changing
direction, multiple jumps, vanishing jumps and jumps changing direction. These
effects are hypothesized to originate from weak links located at the
boundaries of the columnar structure of the NbN films. This hypothesis is
fully consistent with SEM micrographs of these films, and generally agrees
with the extrinsic like behavior of these resonators. To account for the
various nonlinearities observed, a theoretical model assuming local heating of
weak links is suggested. Furthermore, simulation results employing this model
are shown to be in a very good qualitative agreement with measurements.

Such strong sensitive nonlinear effects reported herein may be utilized in the
future in a variety of applications, ranging from qubit coupling in quantum
computation, to signal amplification \cite{IMD amplifier} and to the
demonstration of some important quantum effects in the microwave regime
\cite{nonlinear coupling,Yurke}.

\begin{center}
\textbf{ACKNOWLEDGEMENTS}
\end{center}

E.B. would especially like to thank Michael L. Roukes for supporting the early
stage of this research and for many helpful conversations and invaluable
suggestions. Very helpful conversations with Oded Gottlieb, Gad Koren, Emil
Polturak, and Bernard Yurke are also gratefully acknowledged. This work was
supported by the German Israel Foundation under grant 1-2038.1114.07, the
Israel Science Foundation under grant 1380021, the Deborah Foundation and
Poznanski Foundation.

\bibliographystyle{plain}
\bibliography{apssamp}

\begin{thebibliography}{99}                                                                                               %


\bibitem {squeezing}R. Movshovich, B. Yurke, P. G. Kaminsky, A. D. Smith, A.
H. Silver, R. W. Simon, and M. V. Schneider, Phys. Rev. Lett. \textbf{65},
1419 (1990).

\bibitem {JPA}B. Yurke, P. G. Kaminsky, R. E. Miller, E. A. Whittaker, A. D.
Smith, A. H. Silver, and R. W. Simon, IEEE Trans. Mag. \textbf{25}, 1371 (1989).

\bibitem {Yurke}B. Yurke and E. Buks, quant-ph/0505018.

\bibitem {Eyal}E. Buks and B. Yurke, quant-ph/0511033.

\bibitem {rf driven}I. Siddiqi, R. Vijay, F. Pierre, C. M. Wilson, M.
Metcalfe, C. Riggetti, L. Frunzio, and M. H. Devoret, Phys. Rev. Lett.
\textbf{93}, 207002 (2004).

\bibitem {Direct}I. Siddiqi, R. Vijay, F. Pierre, C. M. Wilson, L. Frunzio, M.
Metcalfe, C. Riggetti, R. J. Schoelkopf, M. H. Devoret, D. Vion and D. Esteve,
Phys. Rev. Lett. \textbf{94}, 027005 (2005).

\bibitem {mixer}P. J. Burke, R. J. Schoelkopf, D. E. Prober, A. Skalare, B. S.
Karasik, M. C. Gaidis, W. R. McGrath, B. Bumble, and H. G. LeDuc, J. Appl.
Phys. \textbf{85}, 1644 (1999).

\bibitem {single photon}R. Sobolewski, A. Verevkin, G. N. Gol'tsman, A.
Lipatov and K. Wilsher, IEEE Trans. Appl. Supercond. \textbf{13}, 1151 (2003).

\bibitem {flux qubit}I. Chiorescu, Y. Nakamura, C. J. P. M. Harmans, and J. E.
Mooij, Science \textbf{299}, 1869 (2003).

\bibitem {Dahm}T. Dahm and D. J. Scalapino, J. Appl. Phys. \textbf{81}, 2002 (1997).

\bibitem {RF power dependence}Z. Ma, E. de Obaldia, G. Hampel, P. Polakos, P.
Mankiewich, B. Batlogg, W. Prusseit, H. Kinder, A. Anderson, D. E. Oates, R.
Ono, and J. Beall, IEEE trans. Appl. Supercond. \textbf{7}, 1911 (1997).

\bibitem {Intrinsic}R. B. Hammond, E. R. Soares, B. A. Willemsen, T. Dahm, D.
J. Scalapino, and J. R. Schrieffer, J. Appl. Phys. \textbf{84}, 5662 (1998).

\bibitem {nonlinear dynamics}C. C. Chen, D. E. Oates, G. Dresselhaus and M. S.
Dresselhaus, Phys. Rev. B. \textbf{45}, 4788 (1992).

\bibitem {Thermally induced nonlinear behaviour}L. F. Cohen, A. L. Cowie, A.
Purnell, N. A. Lindop, S. Thiess, and J. C. Gallop, Supercond. Sci. and
Technol.\textit{ }\textbf{15}, 559 (2002).

\bibitem {microwave nonlinear effects in He-cooled sc microstrip resonators}A.
L. Karuzskii, A. E. Krapivka, A. N. Lykov, A. V. Perestoronin, and A. I.
Golovashkin, Physica B \textbf{329-333}, 1514 (2003).

\bibitem {RF power dependence CP YBCO}Z. Ma, E. D. Obaldia, G. Hampel, P.
Polakos, P. Mankiewich, B. Batlogg, W. Prusseit, H. Kinder, A. Anderson, D. E.
Oates, R. Ono, and J. Beall, IEEE Trans. Appl. Supercond. \textbf{7}, 1911 (1997).

\bibitem {Thermally-induced nonlinearities surface impedance sc YBCO}J. Wosik,
L.-M. Xie, J. H. Miller, Jr., S. A. Long, and K. Nesteruk, IEEE Trans. Appl.
Supercond. \textbf{7}, 1470 (1997).

\bibitem {suspended HTS mw resonator}B. A. Willemsen, J. S. Derov, J. H.
Silva, S. Sridhar, IEEE Trans. Appl. Supercond. \textbf{5}, 1753 (1995).

\bibitem {HTS patch antenna}A. M. Portis, H. Chaloupka, M. Jeck, and A.
Pischke, Superconduct. Sci. \& Technol. \textbf{4}, 436 (1991).

\bibitem {power dependent effects observed for sc stripline resonator}S. J.
Hedges, M. J. Adams, and B. F. Nicholson, Elect. Lett. \textbf{26}, 977 (1990).

\bibitem {mw power handling weak links thermal effects}J. Wosik, L.-M. Xie, R.
Grabovickic, T. Hogan, and S. A. Long,\textit{ }IEEE Trans Appl.
Supercond.\textit{ }\textbf{9}, 2456 (1999).

\bibitem {understanding}D.E. Oates, M. A. Hein, P. J. Hirst, R. G. Humphreys,
G. Koren, and E. Polturak, Physica C. \textbf{372-376}, \ 462 (2002).

\bibitem {Jerusalem}M. A. Golosovsky, H. J. Snortland, and M. R. Beasley,
Phys. Rev. B. \textbf{51}, 6462 (1995).

\bibitem {Yip}S. K. Yip and J. A. Sauls, Phys. Rev. lett. \textbf{69}, 2264 (1992).

\bibitem {vortices}D. E. Oates, H. Xin, G. Dresselhaus, and M. S. Dresselhaus,
IEEE Trans. Appl. Supercond. 11, 2804 (2001).

\bibitem {edge}B. B. Jin and R. X. Wu, J. of Supercond. \textbf{11}, 291 (1998).

\bibitem {anomalies in nonlinear mw surface vs sub effects}A. V. Velichko, D.
W. Huish, M. J. Lancaster, and A. Porch, IEEE Trans. Appl. Supercond.
\textbf{13}, 3598 (2003).

\bibitem {Halbritter}J. Halbritter, J. Appl. Phys.\textit{ }\textbf{68},
\ 6315 (1990).

\bibitem {nonlinear features BB}B. Abdo, E. Segev, Oleg Shtempluck, and E.
Buks, IEEE Trans. Appl. Supercond. (to be published), cond-mat/0501114.

\bibitem {Nonlinear TL}J. H. Oates, R. T. Shin, D. E. Oates, M. J. Tsuk, and
P. P. Nguyen, IEEE Trans. Appl. Supercond. \textbf{3}, 17 (1993).

\bibitem {vortex dynamics}H. Xin, D.E. Oates, G. Dresselhaus, and M. S.
Dresselhaus, J. Supercond. \textbf{14}, 637 (2001).

\bibitem {opposed hammerhead}R. Whiteman, J. Diggins, V. Sch\"{o}llmann, T. D.
Clark, R. J. Prance, H. Prance, and J. F. Ralph, Phys. Lett. A \textbf{234},
\ 205 (1997).

\bibitem {Pinch resonances in rf}H. Prance, T. D. Clark, R. Whiteman, R. J.
Prance, M. Everitt, P. Stiffel, and J. F. Ralph, cond-mat/0411139.

\bibitem {nonlinear multilevel}R. J. Prance, R. Whiteman, T. D. Clark, H.
Prance, V. Sch\"{o}llmann, J. F. Ralph, S. Al-Khawaja, and M. Everitt, Phys.
Rev. Lett. \textbf{82}, 5401 (1999).

\bibitem {Cleland}J. S. Aldridge and A. N. Cleland, cond-mat/0406528.

\bibitem {Kramer}H. A. Kramers, Physica \textbf{7}, 284 (1940).

\bibitem {Sheen}D. M. Sheen, S. M. Ali, D. E. Oates, R. S. Withers, and J. A.
Kong, IEEE Trans. Appl. Supercond. \textbf{1}, 108 (1991).

\bibitem {Bol}M. W. Johnson, A. M. Herr, and A. M. Kadin, J. Appl. Phys.
\textbf{79}, 7069 (1996).

\bibitem {Nonequilibrium}A. M. Kadin and M. W. Johnson, Appl. Phys. Lett.
\textbf{69}, 3938 (1996).

\bibitem {Use}K. Weiser, U. Strom, S. A. Wolf, and D. U. Gubser, J. Appl.
Phys. \textbf{52}, 4888 (1981).

\bibitem {rf superconducting properties of reactively sputtered NbN}S.
Isagawa, J. Appl. Phys. \textbf{52}, 921 (1980).

\bibitem {superconducting properties and NbN structure}Y. M. Shy, L. E. Toth,
and R. Somasundaram, J. Appl. Phys. \textbf{44}, 5539 (1973).

\bibitem {Gardiner}C. W. Gardinar and M. J. Collett, Phys. Rev. A \textbf{31},
3761 (1985).

\bibitem {hot spots}A. VI. Gurevich and R. G. Mints, Rev. Mod. Phys.
\textbf{59}, 941 (1987).

\bibitem {mw hysteretic losses in YBCO and NbN}P. P. Nguyen, D. E. Oates, G.
Dresselhaus, M. S. Dresselhaus, and A. C. Anderson, Phys. Rev. B \textbf{51},
6686 (1995).

\bibitem {Palomba}A. Andreone, A. Cassinese, A. Di Chiara, M. Lavarone, F.
Palomba, A. Ruosi, and R. Vaglio, J. Appl. Phys. \textbf{82}, 1736 (1997).

\bibitem {IMD amplifier}B. Abdo, E. Segev, O. Shtempluck, and E. Buks, cond-mat/0507056.

\bibitem {nonlinear coupling}B. Abdo, E. Segev, O. Shtempluck, and E. Buks, cond-mat/0501236.
\end{thebibliography}

\end{document}